\documentclass[%
 reprint,
superscriptaddress,
 amsmath,amssymb,
 aps,
]{revtex4-2}

\usepackage{graphicx}
\usepackage{dcolumn}
\usepackage{bm}
\usepackage{mathrsfs}
\usepackage{float}

\begin{document}

\preprint{APS/123-QED}

\title{Zeeman and laser site selective spectroscopy of C$_{1}$ point group symmetry Sm$^{3+}$ centers in Y$_{2}$SiO$_{5}$: a parametrized crystal-field analysis for the 4$f^{5}$ configuration}

\author{N. L. Jobbitt}
\affiliation{School of Physical and Chemical Sciences, University of Canterbury, PB4800 Christchurch 8140, New Zealand}
\affiliation{The Dodd-Walls Centre for Photonic and Quantum Technologies, New Zealand}
\author{J.-P. R. Wells}
\email{jon-paul.wells@canterbury.ac.nz}
\author{M. F. Reid}
\email{mike.reid@canterbury.ac.nz}
\affiliation{School of Physical and Chemical Sciences, University of Canterbury, PB4800 Christchurch 8140, New Zealand}
\affiliation{The Dodd-Walls Centre for Photonic and Quantum Technologies, New Zealand}

\date{\today}

\begin{abstract}
Parametrized crystal-field analyses are presented for both the six and seven fold coordinated, C$_{1}$ symmetry Sm$^{3+}$ centers in Y$_{2}$SiO$_{5}$, based on extensive spectroscopic data spanning the infrared to optical regions. Laser site-selective excitation and fluorescence spectroscopy as well as Zeeman absorption spectroscopy performed along multiple crystallographic directions has been utilised, in addition to previously determined $g$ tensors for the $^{6}$H$_{5/2}$Z$_{1}$ and $^{4}$G$_{5/2}$A$_{1}$ states. The resultant analyses give good approximation to the experimental energy levels and magnetic splittings, yielding crystal-field parameters consistent with the few other lanthanide ions for which such analyses are available. 
\end{abstract}

\maketitle

\vspace{2pc}
\noindent{\it Keywords}: Zeeman spectroscopy, laser site selective spectroscopy, low symmetry crystal-field analysis


\section{\label{sec:intro}Introduction}

Rare-earth doped Y$_{2}$SiO$_{5}$ is an attractive option for the development of quantum storage and communication devices due to the low nuclear spin of yttrium ($I=1/2$) and the low abundances of silicon and oxygen isotopes with non-zero nuclear spins \cite{PhysRevA.68.012320, FRAVAL2004347}. Thus the effects of nuclear spin-flips on neighbouring ions on the decoherence times of lanthanide ions doped into this materials system are minimized, resulting in observed hyperfine coherence times exceeding 1 minute for Pr$^{3+}$:Y$_{2}$SiO$_{5}$ and 6 hours in the case of Eu$^{3+}$:Y$_{2}$SiO$_{5}$ \cite{PhysRevLett.111.033601, Zhong}. Y$_{2}$SiO$_{5}$ has therefore been the focus of many studies, providing demonstrations of quantum memories \cite{cit:de_riedmatten, Zhong1392, Ran_i__2017, Zhong, FRAVAL2004347}, quantum gate implementations \cite{PhysRevA.69.032307, PhysRevA.77.022307}, and single photon sources \cite{cit:dibos}. Recently, demonstrations have been made showing control of multiple ions at the single-photon level \cite{Chen592}.

The key technique utilized in obtaining such long coherence times is the zero-first-order Zeeman (ZEFOZ) technique, which involves determining magnetic field points for which the magnetic-hyperfine structure is insensitive to magnetic field fluctuations in any direction. However, ZEFOZ points are notoriously difficult to find experimentally and would optimally be predicted through a crystal-field calculation. Such an approach has the additional benefit of being able to predict the electronic and magnetic-hyperfine structure of the entire 4f$^{N}$ configuration of lanthanide doped Y$_{2}$SiO$_{5}$, whereas more conventional approaches, such as the utilization of a spin Hamiltonian, are only able to predict the magnetic-hyperfine structure of a single state. Previously determined spin Hamiltonian parameters were necessary in obtaining the observed coherence times for Pr$^{3+}$:Y$_{2}$SiO$_{5}$ and Eu$^{3+}$:Y$_{2}$SiO$_{5}$ \cite{PhysRevLett.111.033601, Zhong, cit:Longdell, PhysRevB.85.014429, cit:Longdell2}.

Y$_{2}$SiO$_{5}$ doped with Kramers ions, such as Sm$^{3+}$ and Er$^{3+}$, are appealing candidates in the realization of high-bandwidth quantum storage and communication devices, due to the large hyperfine splittings of such ions, whilst still retaining reasonably long coherence times. Er$^{3+}$:Y$_{2}$SiO$_{5}$ has been observed to have a hyperfine coherence time of 1.3 s at high magnetic field strengths \cite{Ran_i__2017}, while hyperfine coherence times of 1 ms for Yb$^{3+}$:Y$_{2}$SiO$_{5}$ and 1.48 ms for Er$^{3+}$:Y$_{2}$SiO$_{5}$ have been observed without the need of an external magnetic field \cite{cit:ortu, cit:Rakonjac}. These studies utilized previously determined spin Hamiltonian parameters in order to obtain the observed coherence times \cite{PhysRevB.94.155116, PhysRevB.74.214409}. Sm$^{3+}$:Y$_{2}$SiO$_{5}$ serves as a alternative in the realization of high-bandwidth quantum storage and communication devices, due to the small ground state $g$ values, which results in an intrinsic insensitivity to magnetic field fluctuations \cite{cit:jobbitt2}. Furthermore, the 15\% natural abundance of both isotopes with a non-zero nuclear spin ($I = 7/2$ for either $^{147}$Sm or $^{149}$Sm), allows for the possibility of many ZEFOZ points within the complex magnetic-hyperfine structure of Sm$^{3+}$:Y$_{2}$SiO$_{5}$, from which, a single ZEFOZ transition belonging to either isotope could be utilized in the development of high-bandwidth quantum storage and communication devices.

Crystal-field analyses for C$_{1}$ point group symmetry sites, such as those of Y$_{2}$SiO$_{5}$, are difficult to perform as extensive data including orientational information, such as magnetic $g$ values for multiple crystallographic directions, can be required in order to obtain a unique, global fit. Currently, crystal-field analyses have been reported for Ce$^{3+}$, Er$^{3+}$ and Yb$^{3+}$ doped Y$_{2}$SiO$_{5}$ \cite{cit:alizadeh, horvath, cit:Horvath_thesis, cit:zhou}, with further modelling in progress for other ions \cite{Jobbitt2019, cit:mothkuri}.

In this work, we present a parameterized crystal-field analysis for both crystallographic sites of Sm$^{3+}$ doped into Y$_{2}$SiO$_{5}$, which accurately reproduces the electronic and magnetic structure up to the $^{4}$G$_{7/2}$ multiplet at $\sim$20 000 cm$^{-1}$. Laser site-selective excitation and fluorescence, coupled with Zeeman absorption spectroscopy has been performed in order to obtain the data required. Approximately fifty-five electronic energy levels and eighty $g$ values, obtained along all three crystallographic axes, in addition to magnetic splittings determined from the previously determined $g$ tensors of the $^{6}$H$_{5/2}$Z$_{1}$ and $^{4}$G$_{5/2}$A$_{1}$ states \cite{cit:jobbitt2}, were fitted simultaneously for each site in order to uniquely determine the 32 free parameters used in the crystal-field fit. The ground state hyperfine structure has been predicted using the wavefunctions obtained from the crystal-field calculation with similar splittings to what has been reported for Er$^{3+}$:Y$_{2}$SiO$_{5}$ predicted, and significantly larger than what has been observed for Eu$^{3+}$:Y$_{2}$SiO$_{5}$ \cite{horvath, cit:Longdell2}. This, together with the low ground state g-values, suggest Sm$^{3+}$:Y$_{2}$SiO$_{5}$ could be a previously overlooked material for the development of high-bandwidth quantum information storage and communications devices.

\section{\label{sec:experimental}Experimental}

Y$_{2}$SiO$_{5}$ has a monoclinic crystal structure with C$^{6}_{2h}$ space group symmetry. The lattice constants of Y$_{2}$SiO$_{5}$ are; $a = 10.4103$ \AA, $b = 6.7212$ \AA, $c = 12.4905$ \AA, and $ \beta   = 102^{\circ}39'$ \cite{cit:maksimov}. The C$_{2}$ rotation axis corresponds to the $b$ axis, while the $a$ and $c$ axes are located in the plane perpendicular to the $b$ axis. Here we follow the conventions used in \cite{cit:li}, by defining the optical extinction axes $D_{1}$ and $D_{2}$, which are located in the $a$-$c$ plane and are perpendicular to each other in addition to the $b$ axis. In this study we focus on the $X_{2}$ phase of Y$_{2}$SiO$_{5}$, which has two C$_{1}$ symmetry sites, labelled as site 1 and site 2, having coordination numbers of six and seven respectively. Each site possesses two magnetically inequivalent orientations that are related by a 180$^{\circ}$ rotation about the $b$ axis. These two orientations respond differently when a magnetic field is applied outside of the $b$ axis or the $D_{1}$-$D_{2}$ plane \cite{cit:Sun}.

The sample used in this study was grown in the $X_{2}$ phase of Y$_{2}$SiO$_{5}$ via the Czochralski method by Scientific Materials Inc. (Bozeman, USA). The sample was grown with a samarium dopant concentration of 0.5 molar \% and was cut to have dimensions of (5.1 $\pm$ 0.1) mm along the $D_{1}$ axis, (4.9 $\pm$ 0.1) mm along the $D_{2}$ axis, and (6.0 $\pm$ 0.2) mm along the crystallographic $b$ axis.

Laser site-selective excitation and fluorescence spectroscopy was performed using a PTI GL-302 tunable dye laser, which was pumped using a PTI GL-3300 pulsed nitrogen laser. The sample fluorescence was monitored by a Horiba iHR550 spectrometer coupled with either an air cooled Hamamatsu H10330C near infrared photomultiplier tube or a water-cooled Hamamatsu C9727 visible photomultiplier tube connected to appropriate photon counting units. The sample was cooled to 10 K using a Janis CCS-150 closed cycle cryostat coupled to a Lakeshore 325 temperature controller via a resistive heater attached to the back of the sample cold finger. 

Zeeman absorption spectroscopy was performed using a Bruker Vertex-80 Fourier transform infrared (FTIR) spectrometer providing a maximum apodized resolution of 0.075 cm$^{-1}$ and having a spectral range of 700-30,000 cm$^{-1}$, with the entire beam path purged by dry N$_{2}$ gas to minimise atmospheric water. The sample was mounted into the bore of a 4 Tesla simple solenoid and cooled to 4.2~K via thermal contact with the cryostat's liquid helium bath.

\section{\label{sec:results}Results and Discussion}

The 4$f^{5}$ configuration, appropriate to trivalent samarium, consists of 1001 two fold degenerate Kramers states. In this work, we adopt the standard notation of a letter plus a numerical subscript to label the crystal-field levels of the LSJ multiplets. These follow the convention utilised in the standard `Dieke diagram'.

\subsection{\label{sec:temp}Laser site selective spectroscopy}

\begin{figure*}
\centering
	\includegraphics[width=1\textwidth]{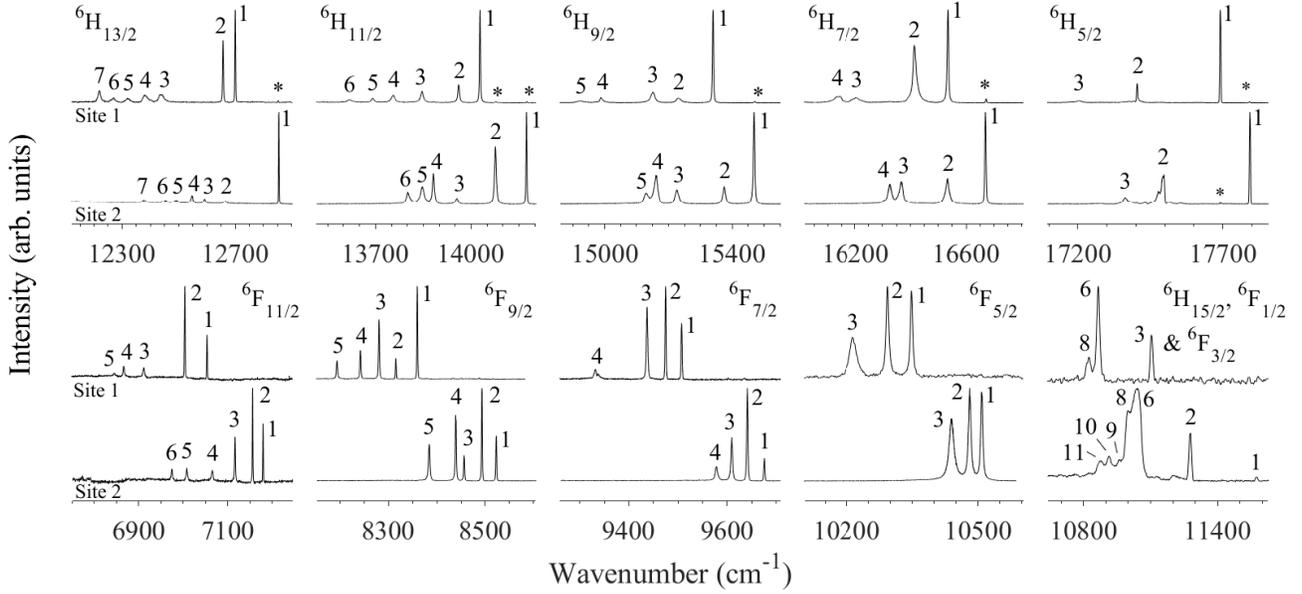}
	\caption{\label{fig1}10 K site-selective fluorescence spectra of site 1 (top spectra in each panel) and site 2 (bottom spectra in each panel) in Sm$^{3+}$:Y$_{2}$SiO$_{5}$. Numbered peaks are assigned to their respective site. Spectral features marked with an `*' are transitions related to the other site. The additional structure observable in the spectra for the $^{6}$H$_{5/2}$, $^{6}$H$_{13/2}$, $^{6}$H$_{15/2}$, $^{6}$F$_{1/2}$ and $^{6}$F$_{3/2}$ multiplets can be attributed to phonon sidebands.}
\end{figure*}

Figure \ref{fig1} shows the 10 K fluorescence spectra for both Sm$^{3+}$ sites in Y$_{2}$SiO$_{5}$ for transitions to all multiplets of the $^{6}$H and $^{6}$F terms. All fluorescence is observed to emanate from the $^{4}$G$_{5/2}$ multiplet regardless of the exact excitation wavelength. To obtain fluorescence to the $^{6}$H$_{5/2}$ multiplet, the $^{6}$H$_{5/2}$Z$_{1}$ $\rightarrow$ $^{4}$F$_{3/2}$B$_{1}$ transition at 18 913 cm$^{-1}$ for site 1 and 18 955 cm$^{-1}$ for site 2 was optically pumped. For all other multiplets, the $^{6}$H$_{5/2}$Z$_{1}$ $\rightarrow$ $^{4}$G$_{5/2}$A$_{1}$ transition at 17 689 cm$^{-1}$ for site 1 and 17 790 cm$^{-1}$ for site 2 was excited. Assignments were confirmed using 10~K absorption spectroscopy.

\begin{figure*}
\centering
	\includegraphics[width=1\textwidth]{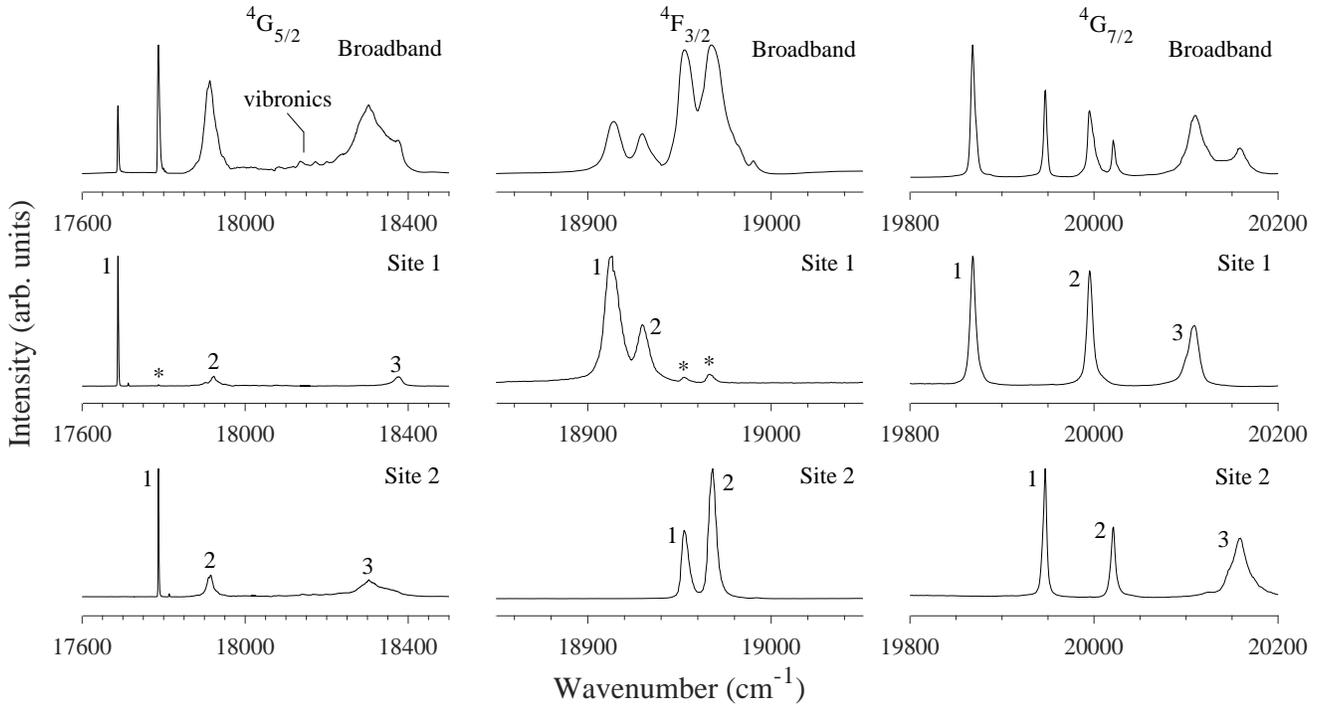}
	\caption{\label{fig2}10 K broadband and site-selective excitation spectra of the $^{4}$G$_{5/2}$, $^{4}$F$_{3/2}$ and $^{4}$G$_{7/2}$ multiplets of both sites in Sm$^{3+}$:Y$_{2}$SiO$_{5}$. Numbered peaks are assignments to their respective site. Spectral features labelled with an `*' are related to site 2. The additional structure observable in the spectra for the $^{4}$G$_{5/2}$ multiplet can be attributed to phonon sidebands.}
\end{figure*}

Figure \ref{fig2} shows 10 K broadband and site-selective excitation spectra for the $^{4}$G$_{5/2}$, $^{4}$F$_{3/2}$ and $^{4}$G$_{7/2}$ multiplets of both sites in Sm$^{3+}$:Y$_{2}$SiO$_{5}$. The broadband spectra were obtained by monitoring all $^{4}$G$_{5/2}$ $\rightarrow$ $^{6}$H$_{9/2}$ fluorescence centered on 15 300 cm$^{-1}$. The $^{4}$G$_{5/2}$ site-selective excitation spectra were obtained by monitoring the $^{4}$G$_{5/2}$A$_{1}$ $\rightarrow$ $^{6}$H$_{9/2}$X$_{1}$ fluorescence at 15 338 cm$^{-1}$ for site 1 and the $^{4}$G$_{5/2}$A$_{1}$ $\rightarrow$ $^{6}$H$_{7/2}$Y$_{1}$ fluorescence at 16 672 cm$^{-1}$ for site 2. Site-selective excitation spectra for the  $^{4}$F$_{3/2}$ and $^{4}$G$_{7/2}$ multiplets were obtained by monitoring the $^{4}$G$_{5/2}$A$_{1}$ $\rightarrow$ $^{6}$H$_{5/2}$Z$_{1}$ fluorescence at 17 689 cm$^{-1}$ and 17 790 cm$^{-1}$ for site 1 and site 2 respectively. The spectral features seen in the site 1 fluorescence spectra while exciting the $^{4}$G$_{5/2}$ and $^{4}$F$_{3/2}$ multiplets are related to site 2 and is due to inter-site energy transfer, as reported previously \cite{cit:jobbitt}.

A total of 56 electronic energy levels are determined for site 1 and 58 electronic energy levels for site 2 through the use of site-selective excitation and fluorescence spectroscopy. These values are summarized in Tables \ref{tab:site1} and \ref{tab:site2} for site 1 and site 2 respectively. 

Phonon sidebands are commonly observed in excitation spectra measured for the Sm$^{3+}$ $^{4}$G$_{5/2}$ multiplet (see for example, refs \cite{cit:wells, cit:wells1, cit:wells2}). For Y$_{2}$SiO$_{5}$ such vibronic structure is also observable as shown in Fig. \ref{fig2}. The observed phonon sideband structure is observed to be shifted from the Sm$^{3+}$ $^{6}$H$_{5/2}\longrightarrow ^{4}$G$_{5/2}$ zero-phonon electronic lines, in the range 100 to 750 cm$^{-1}$. Comparable structure can be observed in some of the fluorescence spectra Fig. \ref{fig1}. These are consistent with the calculated phonon density of states for Y$_{2}$SiO$_{5}$ \cite{cit:luo}.

\subsection{\label{sec:zeeman}Zeeman absorption spectroscopy}

Directional information is required as an input for low symmetry crystal-field analyses in order to obtain a global minimum. Thus the (linear and non-linear) splittings of the Sm$^{3+}$ ion Kramers doublets under the influence of a external magnetic field along multiple crystallographic directions have been measured. Zeeman absorption spectroscopy was performed on all observable transitions between 1500-20000 cm$^{-1}$, using field up to 4~T in magnitude, in order to obtain the required data. Previously, the $g$ tensors of the $^{6}$H$_{5/2}$Z$_{1}$ and $^{4}$G$_{5/2}$A$_{1}$ states for both sites have been reported \cite{cit:jobbitt2}.

\begin{figure}
\centering
	\includegraphics[width=0.5\textwidth]{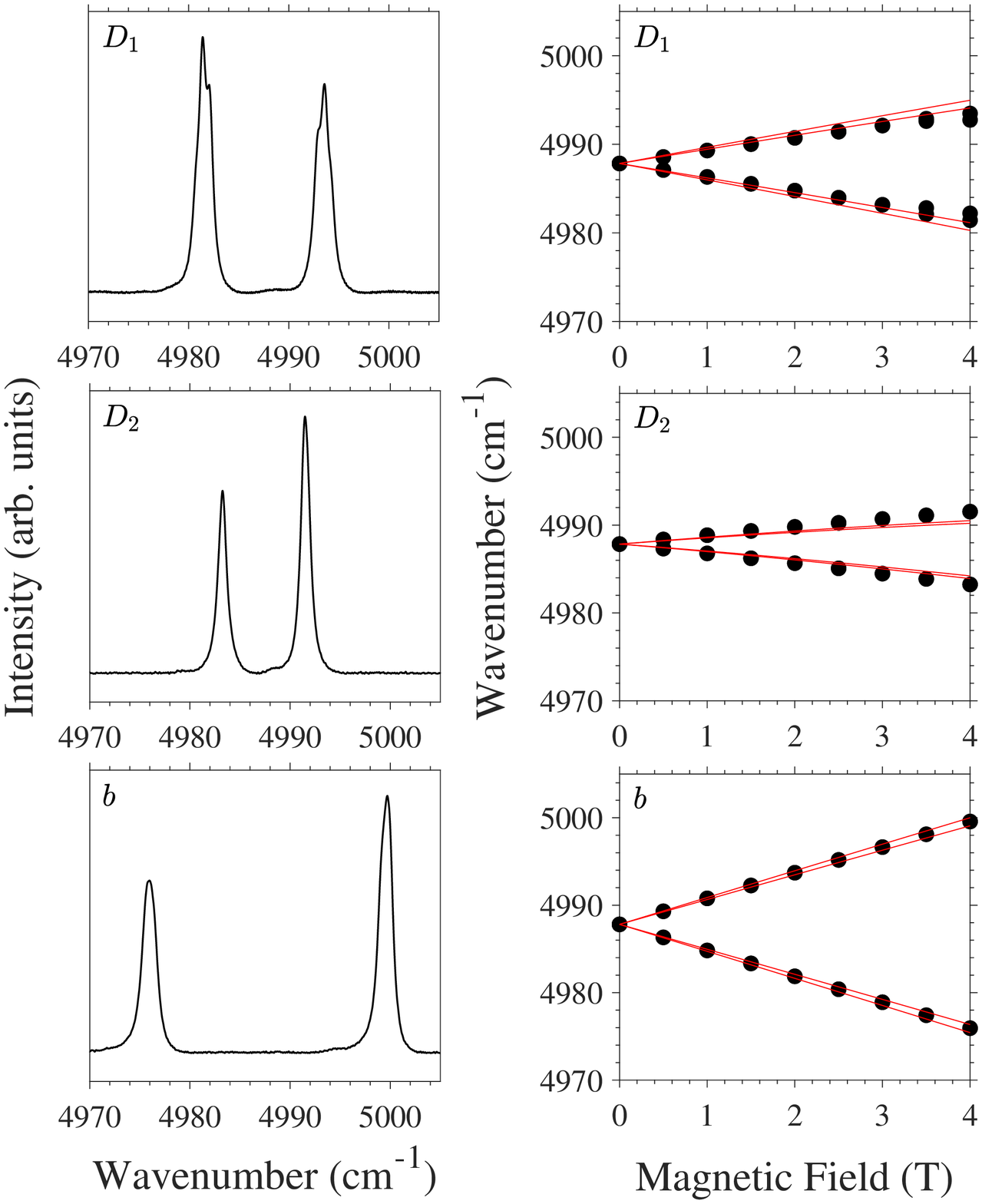}
	\caption{\label{fig3}Magnetic splittings of the $^{6}$H$_{5/2}$Z$_{1}$ $\rightarrow$ $^{6}$H$_{13/2}$V$_{1}$ transition for site 1 in Sm$^{3+}$:Y$_{2}$SiO$_{5}$ with the magnetic field applied along the $D_{1}$ (top), $D_{2}$ (middle) and $b$ (bottom) axes. The left panels show 4.2 K Zeeman absorption spectra at a magnetic field strength of 4 T. The right panels are the experimental splittings, shown by the circles, while the theoretical splittings are given as the red lines.}
\end{figure}

\begin{figure}
\centering
	\includegraphics[width=0.5\textwidth]{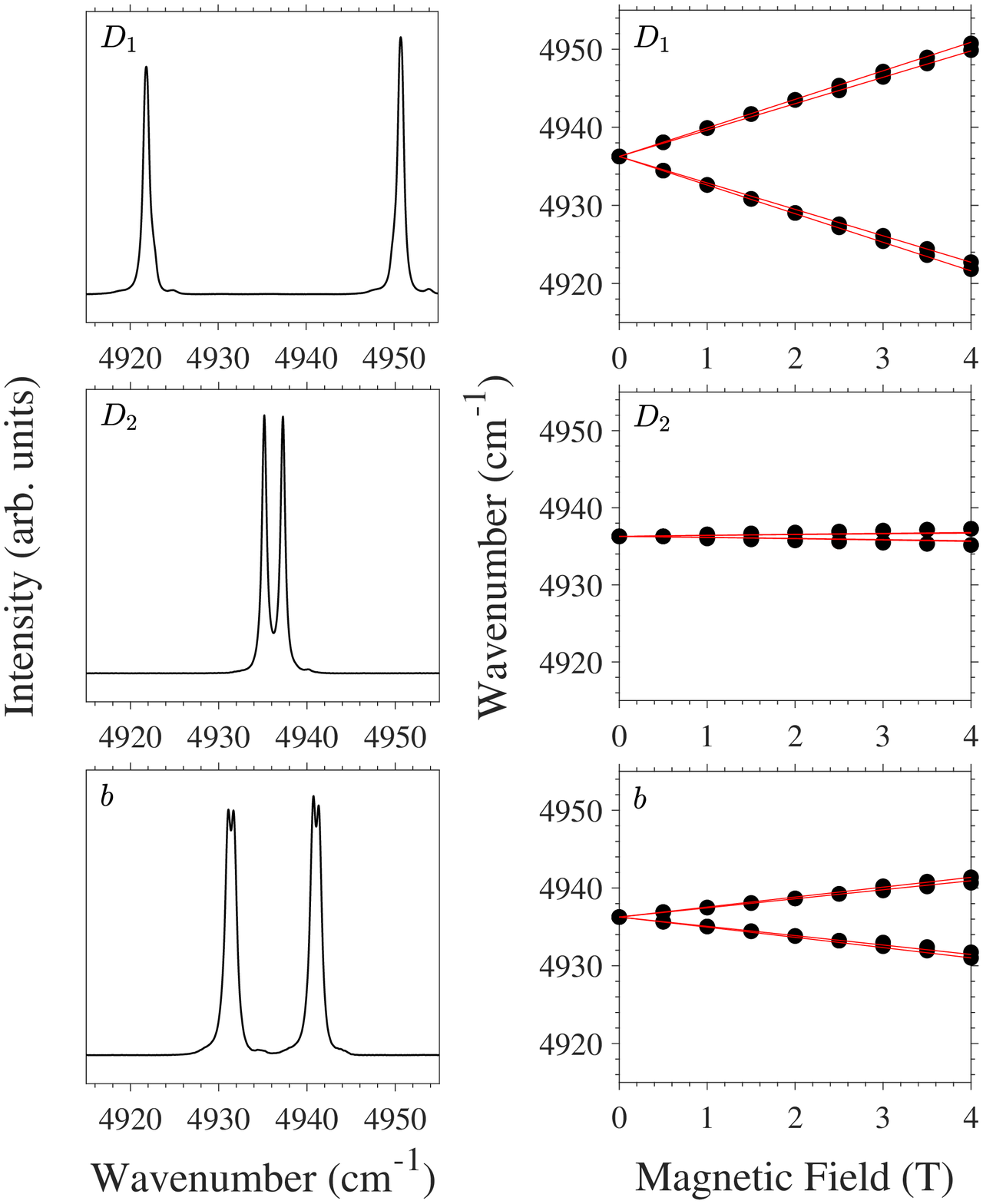}
	\caption{\label{fig4}Magnetic splittings of the $^{6}$H$_{5/2}$Z$_{1}$ $\rightarrow$ $^{6}$H$_{13/2}$V$_{1}$ transition for site 2 in Sm$^{3+}$:Y$_{2}$SiO$_{5}$ with the magnetic field applied along the $D_{1}$ (top), $D_{2}$ (middle) and $b$ (bottom) axes. The left panels show 4.2 K Zeeman absorption spectra at a magnetic field strength of 4 T. The right panels are the experimental splittings, shown by the circles, while the theoretical splittings are given as the red lines.}
\end{figure}

Figures \ref{fig3} and \ref{fig4} shows representative Zeeman absorption spectra for the $^{6}$H$_{5/2}$Z$_{1}$ $\rightarrow$ $^{6}$H$_{13/2}$V$_{1}$ transition for site 1 and site 2 respectively, obtained at 4.2 K. The top, middle and bottom panels shows the magnetic splittings with a magnetic field applied parallel to the $D_{1}$, $D_{2}$ and $b$ axes respectively. The left panels show 4.2 K Zeeman absorption spectra at a magnetic field strength of 4 T. The right panels depicts the experimental magnetic splittings, given by the circles, and the theoretical magnetic splittings, given by the red lines. In general, the calculated values give good approximation to the experimental data.

A total of 72 $g$ values (25 along the $D_{1}$ axis, 22 along the $D_{2}$ axis and 25 along the $b$ axis) for site 1 and 92 $g$ values (34 along the $D_{1}$ axis, 28 along the $D_{2}$ axis and 30 along the $b$ axis) for site 2 were determined utilizing Zeeman absorption spectroscopy. These $g$ values are summarized in Tables \ref{tab:site1} and \ref{tab:site2} for site 1 and site 2 respectively. Due to the very small ground state $g$ values afforded in Sm$^{3+}$:Y$_{2}$SiO$_{5}$ \cite{cit:jobbitt2}, most transitions were found to split into two transitions. In these cases, the observed splitting is approximately the excited state value. 

\subsection{Parameterized crystal-field analysis}

The Hamiltonian used to model the 4f$^{5}$ configuration of Sm$^{3+}$:Y$_{2}$SiO$_{5}$ can be written as \cite{cit:liu}:
\begin{equation}
H = H_{FI} + H_{CF} + H_{HF} + H_{Q} + H_{Z}
\label{eq:hamiltonian}
\end{equation}
where $H_{FI}$ denotes the free-ion interactions, $H_{CF}$ describes the crystal-field interaction, while $H_{Z}$ describes the Zeeman interaction. The free-ion Hamiltonian has an explicit form:
\begin{equation}
\begin{split}
    H_{FI} & =  \ E_{\text{avg}} + \sum_{k = 0,2,4,6}F^{k}f_{k} + \zeta_{nl}A_{SO}(nl)\\ 
    & + \alpha L(L+1) + \beta G(G_{2}) + \gamma G(R_{7})  \\
    & + \sum_{i =2,3,4,6,7,8}T^{i}t_{i} + \sum_{i=0,2,4}M^{i}m_{i} + \sum_{i=2,4,6}P^{i}p_{i}
    \label{eq:FIhamiltonian}
    \end{split}
\end{equation}
where $E_{\text{avg}}$ represents the configuration barycenter, Aspherical electrostatic repulsion is parameterized by the Slater parameters, $F_{k}$, while $\zeta$ describes the spin-orbit interaction. $\alpha$, $\beta$ and $\gamma$ are the two-body (Trees) parameters, while $T_{i}$ are the three-body (Judd) parameters. Finally higher order effects, such as spin-spin and spin-other-orbit in addition to electrostatically correlated magnetic interactions are described by the $M^{i}$ and $P^{i}$ parameters respectively. Here we constrained the $M^{i}$ and $P^{i}$ parameters to the ratios determined by \cite{cit:carnall}.

The crystal-field Hamiltonian is commonly written as: 
\begin{equation}
H_{CF} = \sum_{k,q}B^{k}_{q}C^{(k)}_{q}
\label{eq:CFhamiltonian}
\end{equation}
where the $B^{k}_{q}$ are the coefficients of the expansion of the crystal-field (the crystal-field parameters) in terms of the $C^{(k)}_{q}$ - the Racah spherical tensors, expressed in Wybourne's normalization. In the C$_{1}$ point group symmetry substitutional sites available in the Y$_{2}$SiO$_{5}$ host crystal, $k =2,4,6$ and $q = -k,...,k$. All non-axial ($q\neq 0$) parameters are complex, leading to a total of 27 independent parameters to be determined.

$H_{HF}$ and $H_{Q}$ are the contributions from magnetic dipole and electric quadrupole moment interactions respectively.

The contribution due to the magnetic dipole is given as \cite{cit:judd}:

\begin{equation}
    H_{HF} = a_{l} \sum_{i} \mathbf{N}_{i} \cdot \mathbf{I}
    \label{eq:magnetichyperfinehamiltonian}
\end{equation}

where,

\begin{equation}
    \mathbf{N}_{i} = \mathbf{l}_{i} - \sqrt{10}(s\mathbf{C}^{(2)})_{i}^{(1)}
    \label{eq:N}
\end{equation}

here $\mathbf{l}_{i}$ and $\mathbf{s}_{i}$ are the orbital and spin angular momenta of the $i^{\text{th}}$ electron respectively, while $\mathbf{I}$ is the nuclear spin operator. The magnetic dipole hyperfine parameter, $a_{l}$, is given as \cite{cit:wybourne}:

\begin{equation}
    a_{l} = 2 \mu_{B} g_{n} \mu_{N} \frac{\mu_{0}}{4 \pi} (1 - R)\langle r_{e}^{-3}\rangle
    \label{eq:magnetichyperfineparameter}
\end{equation}

where $\mu_{B}$ and $\mu_{N}$ are the Bohr magneton and nuclear magneton respectively, $\mu_{0}$ is the vacuum permeability, $R$ is a shielding factor describing the effect of the induced closed shell quadrupole moment on the 4f electrons, and $\langle r_{e}^{-3}\rangle$ is the average inverse-cube radius of the 4f orbital.

The contribution due to the electric quadrupole is given as:

\begin{equation}
    H_{Q} = a_{Q}\frac{1}{2}\bigg(\frac{(I+1)(2I+1)(2I+3)}{I(2I-1)}\bigg)^{\frac{1}{2}} \mathbf{U}_{n}^{(2)} \cdot \mathbf{U}_{e}^{(2)}
    \label{eq:electricquadrupolehamiltonian}
\end{equation}

where $\mathbf{U}_{e}^{(2)}$ and $\mathbf{U}_{n}^{(2)}$ are operators for the electronic and nuclear parts of the wave function respectively. The electric quadrupole hyperfine parameter, $a_{Q}$, is given as:

\begin{equation}
    a_{Q} = \frac{-e^{2}Q}{4\pi \epsilon_{0}}(1-R)\langle r_{e}^{-3}\rangle
    \label{eq:electricquadrupoleparameter}
\end{equation}

here $Q$ is the nuclear quadrupole moment. While $a_{l}$ and $a_{Q}$ can both be related to physical constants, they are typically determined from experimental data \cite{cit:Horvath_thesis}.

The final interaction given in Equation (\ref{eq:hamiltonian}), the Zeeman interaction describes the effects of an external magnetic field, $\mathbf{B}$, and can be written as:
\begin{equation}
    H_{Z} = \mu_{B}\sum_{i}^{N}\mathbf{B}\cdot(l_{i}+2s_{i})
    \label{eq:zeemanhamiltonian}
\end{equation}
where $\mu_{B}$ is the Bohr magneton, while $l_{i}$ and $s_{i}$ are the orbital and spin angular momenta of the $i$th electron respectively.

As noted above, in C$_{1}$ symmetry systems such as Y$_{2}$SiO$_{5}$, there are 27 independent crystal-field parameters (counting real and imaginary parts as separate parameters), so crystal-field fits are very challenging. In the absence of directional interactions, such as applied magnetic fields, the energies are invariant under an arbitrary rotation, and so three of these parameters may be set to zero \cite{GNUTEK2008391, doi:10.1080/00268970410001728799}. Different rotations will lead to different parameters being zero and/or different magnitudes. Therefore, comparisons between parameters requires rotations to standardized parameter sets \cite{doi:10.1063/1.449731, RUDOWICZ2016468}.

In this work, magnetic splitting data is used to supply directional information.  We use software developed by Horvath \cite{horvath, PhysRevB.104.155121, cit:Horvath_code} to fit data for low-symmetry systems using electronic energy levels, magnetic splittings, and hyperfine splittings (where available).  A Monte-Carlo algorithm is used to search for the global minimum.  We note that in a C$_{1}$ symmetry environment, measurements with magnetic fields along at least six different axes are required to fully determine the orientation. This is the number of measurements required to determine the g tensor for a Kramers doublet in C$_{1}$ symmetry.

The software developed by Horvath has been applied to extensive data sets for both sites of Er$^{3+}$:Y$_{2}$SiO$_{5}$. Initial work, which included g tensors for two states for each site, as well as hyperfine data \cite{horvath} was later extended by using Zeeman spectroscopy to determine magnetic splittings along the $D_{1}$, $D_{2}$, and $b$ axes of the crystal for dozens of states \cite{PhysRevB.104.155121}.  A similar approach has been independently applied to the two sites in Yb$^{3+}$:Y$_{2}$SiO$_{5}$ \cite{cit:zhou}. 

In the current work, we make use of electronic energy levels, g tensors determined for two electronic states \cite{cit:jobbitt2}, and magnetic splittings along the $D_{1}$, $D_{2}$, and $b$ axes of the crystal for dozens of states. Hyperfine data is not yet available for Sm$^{3+}$:Y$_{2}$SiO$_{5}$.

The spin Hamiltonian for a Kramers doublet may be written as:
\begin{equation}
\mathscr{H} = \mu_{B} \mathbf{B} \cdot \mathbf{g} \cdot \mathbf{S}
\label{eq:spinhamiltonian}
\end{equation}
where $\mathbf{g}$ is the magnetic $g$ tensor, while $\mathbf{S}$ is the vector representation of the electronic spin operator. In our fits we use the experimental g tensors to calculate magnetic splittings at fields sampled from the following parametric spiral:
\begin{equation}
 \mathbf{B} = B_{0}
    \begin{bmatrix}
     \sqrt{1-t^2}\cos(6\pi t) \\
     \sqrt{1-t^2}\sin(6\pi t) \\
     t
    \end{bmatrix},
  \ \ \ t \in [-1,1]
\label{eq:B_vector}
\end{equation}
where $B_{0}$ is the magnitude of the magnetic field vector. Thus, for each site, we are fitting to magnetic splittings in many different directions for two states, in addition to the three axes ($D_{1}$, $D_{2}$, and $b$) for several dozen states. In order to reduce the required computational power, the matrix elements of the crystal-field levels were truncated to 30 000 cm$^{-1}$. The initial crystal-field parameters used in the optimization routine were those found for Er$^{3+}$:Y$_{2}$SiO$_{5}$ \cite{horvath, cit:Horvath_thesis}. All software used to perform the optimization is available from \cite{cit:Horvath_code}. A coarse fit was performed using a basin hopping algorithm, which attempts a random step, followed by a local minimization which utilized the bound optimization by quadratic approximation (BOBYQA) algorithm \cite{cit:wales, cit:Wales_2, cit:johnson}. The Metropolis criterion was applied to each random step, and if accepted, the algorithm was allowed to move to the newly found minima \cite{cit:Metropolis}. A final fit was then conducted using simulated annealing, which allows for the estimation of parameter uncertainties through the use of Markov-chain Monte-Carlo (MCMC) techniques \cite{cit:aster}.

For site 1 a total of 178 experimental data points were fitted simultaneously and are as follows:

\begin{itemize}
    \item 56 electronic energy levels up to the $^{4}$G$_{7/2}$ multiplet at $\sim$20 000 cm$^{-1}$.
    \item 72 $g$ values (25  along  the  $D_{1}$ axis, 22 along the $D_{2}$ axis and 25 along the $b$ axis) corresponding to states up to the $^{4}$G$_{7/2}$ multiplet at $\sim$20 000 cm$^{-1}$.
    \item 25 data points for the magnetic splittings of the $^{6}$H$_{5/2}$Z$_{1}$ state, calculated from the $g$ tensor given in \cite{cit:jobbitt2}, sampled at equally spaced intervals, with $B_{0} = 0.05$ T.
    \item 25 data points for the magnetic splittings of the $^{4}$G$_{5/2}$A$_{1}$ state, calculated from the $g$ tensor given in \cite{cit:jobbitt2}, sampled at equally spaced intervals according to Equation (\ref{eq:B_vector}), with $B_{0} = 0.05$ T.
\end{itemize}

For site 2 a total of 200 experimental data points were fitted simultaneously and are as follows:

\begin{itemize}
    \item 58 electronic energy levels up to the $^{4}$G$_{7/2}$ multiplet at $\sim$20 000 cm$^{-1}$.
    \item 92 $g$ values (34  along  the  $D_{1}$ axis, 28 along the $D_{2}$ axis and 30 along the $b$ axis) corresponding to states up to the $^{4}$G$_{7/2}$ multiplet at $\sim$20 000 cm$^{-1}$.
    \item 25 data points for the magnetic splittings of the $^{6}$H$_{5/2}$Z$_{1}$ state, calculated from the $g$ tensor given in \cite{cit:jobbitt2}, sampled at equally spaced intervals according to Equation (\ref{eq:B_vector}), with $B_{0} = 0.05$ T.
    \item 25 data points for the magnetic splittings of the $^{4}$G$_{5/2}$A$_{1}$ state, calculated from the $g$ tensor given in \cite{cit:jobbitt2}, sampled at equally spaced intervals according to Equation (\ref{eq:B_vector}), with $B_{0} = 0.05$ T.
\end{itemize}

Tables \ref{tab:site1} and \ref{tab:site2} show the calculated and experimental electronic energy levels and $g$ values up to the $^{4}$G$_{7/2}$ multiplet for site 1 and site 2 respectively in Sm$^{3+}$:Y$_{2}$SiO$_{5}$. The standard deviations between theoretical and experimental electronic energy levels were found to be 7.9 cm$^{-1}$ for site 1 and 5.2 cm$^{-1}$ for site 2. Furthermore, our study accurately reproduces the magnetic structure across a large portion of the entire 4f$^{5}$ configuration of Sm$^{3+}$:Y$_{2}$SiO$_{5}$.

\begin{table*}
    \caption{Calculated and experimental electronic energies levels and $g$ values up to $\sim$20 000 cm$^{-1}$ for site 1 in Sm$^{3+}$:Y$_{2}$SiO$_{5}$. All energies are in cm$^{-1}$. Levels marked with a `--' were not observed.} 
    \begin{ruledtabular}
    \begin{tabular}{ ccccrccccccccc }
     & & \multicolumn{3}{c}{Energies}  & &  \multicolumn{8}{c}{$g$ values} \\\cline{3-5}\cline{7-14}
 & & & &  & &  \multicolumn{2}{c}{$D_{1}$ axis}  & & \multicolumn{2}{c}{$D_{2}$ axis} & & \multicolumn{2}{c}{$b$ axis} \\\cline{7-8}\cline{10-11}\cline{13-14}
        Multiplet & State & Calc. & Exp. & Diff.  && Calc. & Exp. && Calc. & Exp. && Calc. & Exp.  \\\hline
        $^{6}$H$_{5/2}$ & Z$_{1}$ & -2   &  0   & $2$      &&  0.47 & 0.36  &&  0.16 & 0.21 &&   0.49  &   0.39 \\
						& Z$_{2}$ & 296  & 287  & $-9$     &&  1.30 &  --   &&  0.36 & --   &&  1.04   &     -- \\
						& Z$_{3}$ & 505  & 486  & $-19$    &&  1.18 &  --   &&  0.41 & --   &&  0.33   &     -- \\
		$^{6}$H$_{7/2}$ & Y$_{1}$ & 1156 & 1154 & $-2$     &&  0.88 &  --   &&  1.00 & --   &&  4.09   &     -- \\
						& Y$_{2}$ & 1278 & 1275 & $-3$     &&  2.06 &  --   &&  3.28 & --   && 1.75    &     -- \\
						& Y$_{3}$ & 1480 & 1484 & $4$      &&  0.87 &  --   &&  3.03 & --   && 2.15    &     -- \\
						& Y$_{4}$ & 1550 & 1551 & $1$      &&  4.31 &  --   &&  0.90 & --   && 2.27    &     -- \\
		$^{6}$H$_{9/2}$ & X$_{1}$ & 2352 & 2351 & $-1$     &&  2.88 &  2.70 &&  1.44 & --   &&  8.04   &   7.76 \\
						& X$_{2}$ & 2453 & 2461 & $8$      &&  5.09 &  --   &&  3.37 & --   &&  3.41   &  -- \\
						& X$_{3}$ & 2539 & 2539 & $0$      &&  2.94 &  --   &&  2.73 & --   &&  2.48   &     -- \\
						& X$_{4}$ & 2697 & 2702 & $5$      &&  1.74 &  --   &&  6.34 & --   &&  3.79   &     -- \\
						& X$_{5}$ & 2761 & 2762 & $1$      &&  7.71 &  --   &&  1.19 & --   &&  3.89   &     -- \\
		$^{6}$H$_{11/2}$& W$_{1}$ & 3656 & 3661 & $5$      &&  4.69 &  4.50 && 1.26  & 2.34 &&  11.41  &  11.58 \\
						& W$_{2}$ & 3723 & 3727 & $4$      &&  6.85 &  --   && 4.40  & --   &&  6.80   &   -- \\
						& W$_{3}$ & 3848 & 3841 & $-7$     &&  4.87 &  --   && 6.67  & --   &&  4.24   &   -- \\
						& W$_{4}$ & 3932 & 3934 & $2$      &&  4.54 &  --   && 3.72  & --   &&  5.12   &   -- \\
						& W$_{5}$ & 3997 & 3995 & $-2$     &&  5.45 &  --   && 8.18  & --   &&  5.84   &     -- \\
						& W$_{6}$ & 4067 & 4071 & $4$      && 11.00 & 11.62 && 1.26  & --   &&  5.34   &     -- \\
		$^{6}$H$_{13/2}$& V$_{1}$ & 4987 & 4988 & $1$      &&  7.41 &  6.05 && 3.36  & 4.44 &&  12.68  &  12.65 \\
						& V$_{2}$ & 5035 & 5031 & $-4$     &&  7.16 &  5.27 && 4.13  & 5.74 &&  10.51  &  11.27 \\
						& V$_{3}$ & 5244 & 5254 & $10$     &&  4.23 &  --   && 3.26  & --   &&  8.91   &   -- \\
						& V$_{4}$ & 5307 & 5312 & $5$      &&  5.09 &  --   && 7.36  & --   &&  4.68   &   -- \\
						& V$_{5}$ & 5364 & 5370 & $6$      &&  8.43 &  --   && 0.98  & --   && 10.14   &   -- \\
						& V$_{6}$ & 5428 & 5423 & $-5$     && 11.00 &  --   && 1.29  & --   &&  5.68   &   -- \\
						& V$_{7}$ & 5471 & 5469 & $-2$     && 13.84 & 11.90 && 1.49  & --   &&  3.72   &     -- \\
		$^{6}$F$_{1/2}$,& S$_{1}$ & 6198  & 6193  & $-5$   &&  9.02 &  7.15 && 4.24  & 6.67  && 13.48   &  13.96 \\
		$^{6}$F$_{3/2}$,& S$_{2}$ & 6394  & 6392  & $-2$   &&  6.66 &  --   && 3.62  & --    && 12.12   &     -- \\
		$^{6}$H$_{15/2}$& S$_{3}$ & 6597  & 6583  & $-14$  &&  1.06 &  1.39 && 1.03  & 1.34  && 0.83    &   1.05 \\
                        & S$_{4}$ & 6635  & --    & --     &&  4.05 &   --  && 9.00  & --    && 6.80    &     -- \\
						& S$_{5}$ & 6695  & --    & --     && 11.57 &   --  && 4.17  & --    && 9.68    &     -- \\
						& S$_{6}$ & 6812  & 6823  & $9$    &&  3.61 &   --  && 4.15  & --    && 3.90    &     -- \\
						& S$_{7}$ & 6839  & --    & --     &&  2.36 &   --  && 3.10  & --    && 2.99    &     -- \\
						& S$_{8}$ & 6862  & 6863  & $1$    &&  1.45 &   --  && 2.16  & --    &&  2.44   &     -- \\
						& S$_{9}$ & 6939  & --    & --     &&  7.10 &   --  && 3.82  & --    &&  9.72   &     -- \\
						& S$_{10}$& 7011  & --    & --     && 14.11 &   --  && 7.66  & --    &&  6.14   &     -- \\
						& S$_{11}$& 7115  & --    & --     && 10.74 &   --  && 14.24 & --    &&  1.24   &     -- \\
		$^{6}$F$_{5/2}$	& R$_{1}$ & 7335  & 7337  & $2$    &&  1.94 &  --   && 2.21  & --    &&  2.07   &   -- \\
						& R$_{2}$ & 7385  & 7392  & $7$    &&  1.65 &  --   && 2.60  & 3.00  &&  1.45   &   -- \\
						& R$_{3}$ & 7479  & 7472  & $-7$   &&  3.13 &  --   && 3.69  & --    &&  1.38   &  -- \\
		$^{6}$F$_{7/2}$	& Q$_{1}$ & 8167  & 8179  & $12$   &&  4.93 &  --   && 3.34  & 3.15  &&  2.31   &   1.66 \\
						& Q$_{2}$ & 8195  & 8212  & $17$   &&  3.66 &  --   && 3.38  & 5.14  &&  2.68   &   2.58 \\
						& Q$_{3}$ & 8246  & 8250  & $4$    &&  3.66 &  3.16 && 2.48  & 2.54  &&  3.75   &   5.97 \\
						& Q$_{4}$ & 8363  & 8353  & $-10$  &&  4.28 &  4.59 && 4.28  & 5.06  &&  4.03   &   3.85 \\
		$^{6}$F$_{9/2}$	& P$_{1}$ & 9315  & 9327  & $12$   &&  5.99 &  7.82 && 6.91  & 4.69  &&  2.79   &   3.91 \\
						& P$_{2}$ & 9366  & 9372  & $6$    &&  4.44 &  6.06 && 4.50  & 4.82  &&  4.51   &   4.19 \\
						& P$_{3}$ & 9400  & 9406  & $6$    &&  4.39 &  2.14 && 3.11  & 3.79  &&  6.37   &   3.04 \\
						& P$_{4}$ & 9432  & 9445  & $13$   &&  3.36 &  3.17 && 1.17  & 2.07  &&  7.33   &   8.48 \\
						& P$_{5}$ & 9495  & 9494  & $-1$   &&  4.81 &  4.12 && 3.61  & 4.82  &&  8.17   &   8.46 \\
		$^{6}$F$_{11/2}$& O$_{1}$ & 10650 & 10635 & $-15$  && 11.90 & 13.17 && 1.30  & --    &&  6.95   &   6.46 \\
						& O$_{2}$ & 10683 & 10685 & $2$    &&  2.80 & 1.37  && 11.42 & 13.14 &&  4.77   &   4.53 \\
						& O$_{3}$ & 10772 & 10777 & $5$    &&  5.80 & 7.81  && 6.33  & 5.52  &&  5.03   &   3.90 \\
						& O$_{4}$ & 10837 & 10822 & $-15$  &&  6.54 & 8.49  && 7.80  & 9.28  &&  2.90   &   2.04 \\
						& O$_{5}$ & 10862 & 10844 & $-18$  &&  7.66 & 6.17  && 7.77  & 5.83  &&  2.68   &   7.31 \\
						& O$_{6}$ & 10927 & 10919 & $-8$   &&  1.62 & 3.05  && 0.65  & --    &&  14.79  &  14.84 \\
		$^{4}$G$_{5/2}$	& A$_{1}$ & 17690 & 17689 & $-1$   &&  1.19 & 1.07  && 1.09  & 1.29  &&  2.43   &   2.46 \\
						& A$_{2}$ & 17920 & 17922 & $2$    &&  1.00 &    -- && 2.47  & --    &&  0.72   &     -- \\
						& A$_{3}$ & 18350 & 18355 & $5$    &&  1.95 &    -- && 1.15  & --    &&  1.74   &     -- \\
		$^{4}$F$_{3/2}$	& B$_{1}$ & 18912 & 18913 & $1$    &&  0.76 &    -- && 0.97  & --    &&  0.53   &     -- \\
						& B$_{2}$ & 18927 & 18929 & $2$    &&  0.94 &    -- && 0.45  & --    &&  0.73   &     -- \\
		$^{4}$G$_{7/2}$	& C$_{1}$ & 19866 & 19868 & $2$    &&  2.07 & 2.12  && 0.85  & --    &&  5.28   &   6.28 \\
						& C$_{2}$ & 20002 & 19995 & $-7$   &&  2.43 & 3.40  && 4.96  & 3.58  &&  0.94   &   1.98 \\
						& C$_{3}$ & 20111 & 20109 & $-2$   &&  1.95 &    -- && 2.50  & --    &&  2.45   &     -- \\
						& C$_{4}$ & 20258 & --    & --     &&  8.23 &    -- && 1.82  & --    &&  5.63   &     -- \\
    \end{tabular}
    \end{ruledtabular}
    \label{tab:site1}
\end{table*}

\begin{table*}
    \caption{Calculated and experimental electronic energies levels and $g$ values up to $\sim$20 000 cm$^{-1}$ for site 2 in Sm$^{3+}$:Y$_{2}$SiO$_{5}$. All energies are in cm$^{-1}$. Levels marked with a `--' were not observed.} 
    \begin{ruledtabular}
    \begin{tabular}{ ccccrccccccccc }
     & & \multicolumn{3}{c}{Energies}  & &  \multicolumn{8}{c}{$g$ values} \\\cline{3-5}\cline{7-14}
 & & & &  & &  \multicolumn{2}{c}{$D_{1}$ axis}  & & \multicolumn{2}{c}{$D_{2}$ axis} & & \multicolumn{2}{c}{$b$ axis} \\\cline{7-8}\cline{10-11}\cline{13-14}
        Multiplet & State & Calc. & Exp. & Diff.  && Calc. & Exp. && Calc. & Exp. && Calc. & Exp.  \\\hline
        $^{6}$H$_{5/2}$ & Z$_{1}$ & 1    &  0   & $-1$    &&  0.59 & 0.52  &&  0.05 & 0.08  &&  0.24   &   0.15 \\
						& Z$_{2}$ & 301  & 300  & $-1$    &&  0.69 &  --   &&  0.95 & --    &&  0.44   &     -- \\
						& Z$_{3}$ & 438  & 429  & $-9$    &&  0.27 &  --   &&  0.75 & --    &&  1.41   &     -- \\
		$^{6}$H$_{7/2}$ & Y$_{1}$ & 1121 & 1119 & $-2$    &&  4.78 &  --   &&  1.18 & --    &&  0.91   &     -- \\
						& Y$_{2}$ & 1258 & 1256 & $-2$    &&  3.66 &  --   &&  1.28 & --    &&  1.99   &     -- \\
						& Y$_{3}$ & 1419 & 1421 & $2$     &&  2.51 &  --   &&  3.45 & --    &&  1.92   &     -- \\
						& Y$_{4}$ & 1465 & 1463 & $-2$    &&  1.27 &  --   &&  2.48 & --    &&  3.23   &     -- \\
		$^{6}$H$_{9/2}$ & X$_{1}$ & 2317 & 2322 & $5$     &&  8.41 &  7.92 && 1.95  & 1.48  &&  1.15   &   1.63 \\
						& X$_{2}$ & 2412 & 2415 & $3$     &&  6.74 &  6.15 && 1.12  & --    &&  3.66   &   4.32 \\
						& X$_{3}$ & 2558 & 2564 & $6$     &&  4.12 &  --   && 3.69  & --    &&  2.10   &     -- \\
						& X$_{4}$ & 2626 & 2627 & $1$     &&  3.28 &  --   && 5.46  & --    &&  5.41   &   6.49 \\
						& X$_{5}$ & 2661 & 2660 & $-1$    &&  2.54 &  --   && 4.39  & 6.07  &&  4.40   &     -- \\
		$^{6}$H$_{11/2}$& W$_{1}$ & 3610 & 3615 & $5$     && 11.92 & 11.93 && 1.41  & 0.69  &&  2.75   &   2.56 \\
						& W$_{2}$ & 3708 & 3713 & $5$     &&  9.97 & 10.10 && 1.11  & --    &&  3.52   &   4.00 \\
						& W$_{3}$ & 3837 & 3833 & $-4$    &&  7.25 & 7.06  && 3.34  & --    &&  3.14   &   --   \\
						& W$_{4}$ & 3904 & 3906 & $2$     &&  2.85 & 3.90  && 6.33  & 4.10  &&  2.93   &   --   \\
						& W$_{5}$ & 3937 & 3940 & $3$     &&  5.99 &  --   && 2.60  & --    &&  7.83   &     -- \\
						& W$_{6}$ & 3985 & 3987 & $2$     &&  1.59 & --    && 5.35  & 8.02  &&  8.75   &   7.99 \\
		$^{6}$H$_{13/2}$& V$_{1}$ & 4935 & 4936 & $1$     && 15.09 & 15.01 && 0.58  & 1.12  &&  5.31   &   5.17 \\
						& V$_{2}$ & 5125 & 5124 & $-1$    && 13.44 & 13.89 && 2.83  & --    &&  2.85   &   --   \\
						& V$_{3}$ & 5201 & 5199 & $-2$    &&  8.84 & 9.01  && 5.89  & 6.97  &&  4.58   &   3.17 \\
						& V$_{4}$ & 5242 & 5244 & $2$     &&  4.14 & 3.12  && 8.36  & 10.04 &&  4.20   &   --   \\
						& V$_{5}$ & 5304 & 5301 & $-3$    &&  5.80 & 5.68  && 7.41  & --    &&  4.75   &   --   \\
						& V$_{6}$ & 5342 & 5337 & $-5$    &&  2.99 & 8.23  && 7.05  & 7.23  &&  9.76   &   --   \\
						& V$_{7}$ & 5409 & 5414 & $5$     &&  2.60 & --    && 6.39  & 8.76  &&  13.90  &   12.35 \\
		$^{6}$F$_{1/2}$,& S$_{1}$ & 6218 & 6217  & $-1$   && 17.87 & 17.86 &&  1.24 & 0.95  && 6.30    &  6.70 \\
		$^{6}$F$_{3/2}$,& S$_{2}$ & 6528 & 6514  & $-14$  &&  1.07 &   --  &&  1.92 & 0.99  &&  1.44   &   1.68 \\
		$^{6}$H$_{15/2}$& S$_{3}$ & 6546 & --    & --     && 10.15 &   --  &&  4.66 & --    &&  3.78   &     -- \\
                        & S$_{4}$ & 6582 & --    & --     &&  5.01 &   --  && 10.48 & --    &&   2.97  &     -- \\
						& S$_{5}$ & 6665 & --    & --     &&  5.84 &   --  &&  6.97 & --    &&   6.19  &     -- \\
						& S$_{6}$ & 6748 & 6752  & $4$    &&  1.67 &   --  &&  3.76 & --    &&   6.43  &     -- \\
						& S$_{7}$ & 6772 & --    & --     &&  2.36 &   --  &&  3.10 & --    &&   2.93  &     -- \\
						& S$_{8}$ & 6798 & 6792  & $-6$   &&  1.65 &   --  &&  2.09 & --    &&   1.15  &     -- \\
						& S$_{9}$ & 6838 & 6832  & $-6$   &&  2.63 &   --  &&  4.09 & --    &&  10.64  &     -- \\
						& S$_{10}$& 6875 & 6873  & $-2$   && 10.99 &   --  && 10.40 & --    &&   4.16  &     -- \\
						& S$_{11}$& 6917 & 6912  & $-5$   &&  3.78 &   --  &&  3.81 & --    &&  14.38  &     -- \\
		$^{6}$F$_{5/2}$	& R$_{1}$ & 7280 & 7281  & $1$    &&  4.14 & 3.75  &&  2.72 & --    &&  0.55   &   2.54 \\
						& R$_{2}$ & 7303 & 7308  & $5$    &&  2.28 & 3.27  &&  0.69 & --    && 2.94    &   2.38 \\
						& R$_{3}$ & 7356 & 7350  & $-6$   &&  2.64 &  --   &&  2.98 & --    &&  2.56   &     -- \\
		$^{6}$F$_{7/2}$	& Q$_{1}$ & 8107 & 8117  & $10$   &&  2.93 & 2.26  &&  4.01 & 1.55  &&  3.00   &   2.50 \\
						& Q$_{2}$ & 8143 & 8152  & $9$    &&  2.46 & 2.67  &&  4.82 & 3.44  &&  3.44   &   4.29 \\
						& Q$_{3}$ & 8185 & 8184  & $-1$   &&  6.91 & 7.16  && 2.17  & 1.88  &&  2.34   &   2.42 \\
						& Q$_{4}$ & 8216 & 8214  & $-2$   &&  1.67 & --    && 5.58  & 4.64  &&  4.08   &   5.61 \\
		$^{6}$F$_{9/2}$	& P$_{1}$ & 9269 & 9270  & $1$    &&  5.35 &  2.80 && 4.07  & 6.40  &&  7.83   &   7.24 \\
						& P$_{2}$ & 9297 & 9300  & $3$    &&  6.29 &  3.39 && 1.85  & 1.86  &&  4.80   &   6.03 \\
						& P$_{3}$ & 9324 & 9337  & $13$   &&  4.78 &  3.38 && 7.32  & 5.32  &&  2.76   &   2.59 \\
						& P$_{4}$ & 9346 & 9354  & $8$    &&  4.25 &  8.55 && 6.04  & --    &&  3.31   &   2.79 \\
						& P$_{5}$ & 9405 & 9408  & $3$    && 10.33 & 11.18 && 1.86  & --    &&  3.22   &  --   \\
		$^{6}$F$_{11/2}$& O$_{1}$ & 10623 & 10612 & $-11$ &&  4.05 &  4.83 && 6.20  & 6.57  && 10.97   &  11.93 \\
						& O$_{2}$ & 10645 & 10635 & $-10$ &&  4.02 & 2.69  && 2.94  & 2.70  &&  6.59   &   8.66 \\
						& O$_{3}$ & 10683 & 10675 & $-8$  &&  6.06 & 5.72  && 4.01  & 4.40  &&  3.10   &   4.32 \\
						& O$_{4}$ & 10729 & 10726 & $-3$  &&  9.49 & 9.35  &&  2.42 & 2.48  &&  4.53   &   4.39 \\
						& O$_{5}$ & 10774 & 10783 & $9$   && 11.47 & 10.07 &&  6.15 &   --  &&  7.90   &   --   \\
						& O$_{6}$ & 10820 & 10817 & $-3$  && 13.25 & 14.24 &&  8.49 &   --  &&  0.83   &   --   \\
		$^{4}$G$_{5/2}$	& A$_{1}$ & 17795 & 17790 & $-5$  &&  2.97 & 3.29  &&  0.37 &  0.38 &&  0.78   &   0.77 \\
						& A$_{2}$ & 17922 & 17915 & $-7$  &&  2.31 &    -- &&  0.59 &   --  &&   1.01  &     -- \\
						& A$_{3}$ & 18282 & 18291 & $9$   &&  1.15 &    -- &&  2.18 &   --  &&   1.69  &     -- \\
		$^{4}$F$_{3/2}$	& B$_{1}$ & 18953 & 18955 & $2$   &&  0.33 &  0.79 &&  1.17 & 0.75  &&    0.69 &     -- \\
						& B$_{2}$ & 18971 & 18968 & $-3$  &&  0.88 &    -- &&  0.39 & 0.91  &&    0.99 &   0.79 \\
		$^{4}$G$_{7/2}$	& C$_{1}$ & 19947 & 19947 & $0$   &&  6.16 &  6.99 &&  2.25 & 1.21  &&    0.82 &   1.07 \\
						& C$_{2}$ & 20016 & 20021 & $5$   &&  3.99 &  4.75 &&  2.06 &   --  &&    2.80 &   2.25 \\
						& C$_{3}$ & 20159 & 20159 & $0$   &&  3.72 &    -- &&  1.43 &   --  &&    0.74 &     -- \\
						& C$_{4}$ & 20296 & --    & --    &&  1.38 &    -- &&  4.02 &   --  &&    3.51 &     -- \\
    \end{tabular}
    \end{ruledtabular}
    \label{tab:site2}
\end{table*}

Table \ref{tab:Sm_Parameters} shows the fitted free-ion and crystal-field parameters for both sites of Sm$^{3+}$:Y$_{2}$SiO$_{5}$, in the  ($D_{1}$, $D_{2}$, $b$) frame. The two- and three-body interactions, in addition to the higher order effects given in Equation (\ref{eq:FIhamiltonian}) were held fixed to the values found for Sm$^{3+}$:LaF$_{3}$ and are shown in Table \ref{tab:Fixed_Parameters} \cite{cit:carnall}. Note that the two-fold symmetry of the crystal (not the sites) about the $b$ axis gives rise to two magnetically inequivalent orientations for each site, which in general have different magnetic splittings unless the field is oriented along the $b$ axis, or in the  $D_{1}$-$D_{2}$ plane. These parameter sets are related by a two-fold rotation, which gives a phase $e^{i m \pi}$. This has the result of multiplying crystal-field parameters with odd $q$ by -1.

The parameter uncertainties were estimated through the use of MCMC techniques by sampling the posterior probability distribution. A total of 5 000 000 trials were attempted for both sites with 423 818 accepted steps for site 1 and 401 830 accepted steps for site 2. This aligns with the Metropolis algorithms $\sim$10 \% acceptance rate recommended for this technique, which was fine tuned through altering the step size in the optimization routine \cite{cit:aster}. The algorithm was allowed to `burn in' (become confined within the area of the global minimum), and every 10$^{\text{th}}$ element of the remaining 50 000 steps were then selected to ensure that the samples were not correlated. Our Monte-Carlo approach is designed to locate the global minimum, and trials with different starting parameters lead to similar fitted parameters.

\begin{table*}
	\caption{Fitted values for the free-ion and crystal-field parameters and their related uncertainties of site 1 and site 2 in Sm$^{3+}$:Y$_{2}$SiO$_{5}$. All values are in cm$^{-1}$. Note that the two magnetically inequivalent orientations for each site, are related by a two-fold rotation about the $b$ axis. This has the result of multiplying crystal-field parameters with odd q by -1.} 
	\begin{ruledtabular}
	\begin{tabular}{ cccccc }
	\rule{0pt}{10pt}	          & \multicolumn{2}{c}{Site 1} && \multicolumn{2}{c}{Site 2} \\\cline{2-3}\cline{5-6}   
	\rule{0pt}{10pt}	Parameter & Site 1 & Site 1 Uncertainty && Site 2 & Site 2 Uncertainty \\\hline
	\rule{0pt}{10pt} $E_{\text{avg}}$  & $47316$ & $1$ && $47338$  &  $1$ \\
    $F^{2}$       & $78297$      & $3$      && $78391$          &  $3$ \\
	$F^{4}$       & $56463$      & $5$      && $56606$          &  $5$ \\
	$F^{6}$       & $39727$      & $4$      && $39876$          &  $6$ \\
	$\zeta$       & $1165.98$    & $1$      && $1169$           & $1$ \\
	$B^{2}_{0}$   & $-511$       & $5$      && $446$            & $9$ \\
	$B^{2}_{1}$   & $603+241i$   & $6 + 5i$ && $329-104i$       & $8 + 8i$ \\
	$B^{2}_{2}$   & $91-62i$     & $5 + 4i$ && $-460-62i$       & $7 + 8i$ \\
	$B^{4}_{0}$   & $2068$       & $7$      && $810$            & $8$ \\
	$B^{4}_{1}$   & $167+507i$   & $5 + 13i$&& $-539+40i$       & $10 + 8i$ \\
	$B^{4}_{2}$   & $-144+466i$  & $6 + 7i$ && $-85-605i$       & $8 + 11i$ \\
	$B^{4}_{3}$   & $-299-112i$  & $4 + 2i$ && $-181+440i$      & $9 + 10i$ \\
	$B^{4}_{4}$   & $-901+1021i$ & $6 + 3i$ && $-994+80i$       & $10 +9i$ \\
	$B^{6}_{0}$   & $-45$        & $5$      && $521$            & $9$ \\
	$B^{6}_{1}$   & $534-52i$    & $3 + 8i$ && $-241-68i$       & $11 + 5i$ \\
	$B^{6}_{2}$   & $109+271i$   & $4 + 2i$ && $-324+161i$      & $12 + 9i$ \\
	$B^{6}_{3}$   & $266+84i$    & $5 + 6i$ && $-54+22i$        & $7 + 7i$ \\
	$B^{6}_{4}$   & $371-199i$   & $6 + 5i$ && $-17+44i$        & $8 + 10i$ \\
	$B^{6}_{5}$   & $-155-27i$   & $12 + 8i$&& $38+20i$         & $8 + 14i$ \\
	$B^{6}_{6}$   & $-81-46i$    & $5 + 7i$ && $423+253i$       & $6 + 7i$ \\
	\end{tabular}
	\end{ruledtabular}
	\label{tab:Sm_Parameters}
\end{table*}

\begin{table}
	\caption{Parameters that were held fixed during the fitting process and were set to those found by Carnall \textit{et al.} in Sm$^{3+}$:LaF$_{3}$ \cite{cit:carnall}.} 
	\begin{ruledtabular}
	\begin{tabular}{ cc }      
    Parameter     & Value\\\hline
	$\alpha$      & $20.16$\\
	$\beta$       & $-566.9$\\
	$\gamma$      & $1500$\\
	$T^{2}$       & $300$\\
	$T^{3}$       & $36$\\
	$T^{4}$       & $56$\\
	$T^{6}$       & $-347$\\
	$T^{7}$       & $373$\\
	$T^{8}$       & $348$\\
	$M^{0}$       & $2.6$\\
	$P^{2}$       & $357$\\
	\end{tabular}
	\end{ruledtabular}
	\label{tab:Fixed_Parameters}
\end{table}

Table \ref{tab:gtensor} shows the $g$ tensors determined from our crystal-field analysis for the $^{6}$H$_{5/2}$Z$_{1}$ and $^{4}$G$_{5/2}$A$_{1}$ states for both sites in Sm$^{3+}$:Y$_{2}$SiO$_{5}$. The experimental $g$ tensors are also given for comparison \cite{cit:jobbitt2}. It can be seen that the calculation gives good account of the experimental data. 
\begin{table*}
\centering
	\caption{Calculated and experimental \cite{cit:jobbitt2} $g$ tensors for the $^{6}$H$_{5/2}$Z$_{1}$ and $^{4}$G$_{5/2}$A$_{1}$ states for both sites in Sm$^{3+}$:Y$_{2}$SiO$_{5}$.} 
	\begin{ruledtabular}
	\begin{tabular}{ clccccccccccccc }
	&& \multicolumn{6}{c}{Site 1} && \multicolumn{6}{c}{Site 2} \\\cline{3-8}\cline{10-15} 
	State  &  &   $g_{xx}$ & $g_{yy}$ & $g_{zz}$ & $g_{xy}$ & $g_{xz}$ & $g_{yz}$  &&   $g_{xx}$ & $g_{yy}$ & $g_{zz}$ & $g_{xy}$ & $g_{xz}$ & $g_{yz}$ \\\hline
	$^{6}$H$_{5/2}$Z$_{1}$ & Calc.    & $0.448$    & $0.084$     & $0.487$     & $-0.120$    & $0.044$     & $0.068$  &&  $0.574$    & $0.047$     & $0.186$     & $0.010$    & $-0.147$     & $-0.009$  \\
	                        & Expt.  & $0.351$    & $0.209$     & $0.382$     & $-0.016$    & $0.078$     & $-0.007$ && $0.512$     & $0.067$     & $0.135$     & $-0.040$    & $-0.054$     & $0.009$ \\
	$^{4}$G$_{5/2}$A$_{1}$ & Calc.    & $1.113$    & $0.979$     & $2.405$     & $-0.373$    & $-0.203$    & $0.265$ &&  $2.932$    & $0.260$     & $0.531$     & $0.184$    & $-0.526$     & $-0.112$ \\
	                        & Expt.  & $1.025$    & $1.248$     & $2.446$     & $-0.257$    & $0.166$     & $0.202$ && $3.264$     & $0.183$     & $0.714$     & $-0.311$    & $0.262$      & $0.125$ \\
	\end{tabular}
	\end{ruledtabular}
	\label{tab:gtensor}
\end{table*}

Table \ref{tab:Sm_hyperfine} shows the predicted zero-field hyperfine splittings of the $^{6}$H$_{5/2}$Z$_{1}$ ground and $^{4}$G$_{5/2}$A$_{1}$ excited states for both sites in Sm$^{3+}$:Y$_{2}$SiO$_{5}$.  As the hyperfine parameters are largely host invariant, the hyperfine parameter determined for Sm$^{3+}$:CaF$_{2}$, co-doped with Na$^{3+}$ ions, was used in our predictions \cite{Horvath_2018}. It should be noted that the nuclear-quadrupole parameter was not determined in \cite{Horvath_2018} and was set to zero in our predictions. The ground state ($^{6}$H$_{5/2}$Z$_{1}$) hyperfine structure is predicted to span about 4 GHz for both sites, while the excited state ($^{4}$G$_{5/2}$A$_{1}$) of both sites is expected to span about 2.5 GHz. These are splittings on the same order as that of Er$^{3+}$:Y$_{2}$SiO$_{5}$, and are much larger that the hyperfine splittings of Eu$^{3+}$:Y$_{2}$SiO$_{5}$. As such we believe Sm$^{3+}$:Y$_{2}$SiO$_{5}$ could be an attractive candidate for high-bandwidth quantum memories \cite{horvath, cit:Longdell2}.

\begin{table}[h!]
\centering
	\caption[Calculated zero-field hyperfine splittings of the Z$_{1}$ and A$_{1}$ states]{Calculated zero-field hyperfine splittings of the $^{6}$H$_{5/2}$Z$_{1}$ and $^{4}$G$_{5/2}$A$_{1}$ states for both sites in Sm$^{3+}$:Y$_{2}$SiO$_{5}$. Level 1 is defined as the zero point for both states. All values are in GHz.} 
	\begin{ruledtabular}
	\begin{tabular}{ cccccc }
            &  \multicolumn{2}{c}{Site 1} && \multicolumn{2}{c}{Site 2}\\\cline{2-3}\cline{5-6}
	Level  & $^{6}$H$_{5/2}$Z$_{1}$ &  $^{4}$G$_{5/2}$A$_{1}$ && $^{6}$H$_{5/2}$Z$_{1}$ &  $^{4}$G$_{5/2}$A$_{1}$  \\\hline
	1  & 0         & 0          && 0           & 0         \\
	2  & 0         & 0.00009    && 0           & 0         \\
    3  & 0.54397   & 0.19639    && 0.62839     & 0.37123   \\
	4  & 0.54418   & 0.20260    && 0.62839     & 0.37123   \\
	5  & 1.05596   & 0.32120    && 1.25634     & 0.73821   \\
	6  & 1.07575   & 0.38388    && 1.25640     & 0.73959   \\
	7  & 1.40441   & 0.41584    && 1.85338     & 1.05146   \\
	8  & 1.93090   & 1.64454    && 1.91618     & 1.15621   \\
	9  & 2.12520   & 1.65656    && 2.46357     & 1.36463   \\
	10 & 2.52932   & 1.82643    && 2.56955     & 1.65387   \\
	11 & 2.88649   & 1.90452    && 3.14710     & 1.90710   \\
	12 & 2.90277   & 1.94182    && 3.14722     & 1.91052   \\
	13 & 3.42318   & 2.11417    && 3.77467     & 2.26925   \\
	14 & 3.42336   & 2.11635    && 3.77467     & 2.26925   \\
	15 & 3.96940   & 2.34102    && 4.40302     & 2.63865   \\
	16 & 3.96940   & 2.34105    && 4.40302     & 2.63865   \\
	\end{tabular}
	\end{ruledtabular}
	\label{tab:Sm_hyperfine}
\end{table}

Crystal-field parameters in higher-symmetry sites show a reasonably consistent trend across the rare-earth series (see, for example \cite{cit:carnall}). For Y$_{2}$SiO$_{5}$ the only published fits that make use of directional magnetic data are for for Er$^{3+}$ \cite{horvath, PhysRevB.104.155121}, Yb$^{3+}$ \cite{cit:zhou}, and this work. In comparing parameter sets for different ions, we must take into account that for each site there are two magnetically-inequivalent orientations, which are related by changing the sign of the odd-$q$ crystal-field parameters.  We note that for Yb$^{3+}$ there are only a small number of electronic levels and for Sm$^{3+}$, the hyperfine splittings are not available. The parameters obtained here are broadly similar to the parameters for Er$^{3+}$, but it is an open question how much they would change with additional data in the fit.

\section{Conclusions}

We have presented parameterized crystal-field analyses for both the six and seven fold coordinated C$_{1}$ point group symmetry substitutional sites of Sm$^{3+}$:Y$_{2}$SiO$_{5}$, based upon a large quantity of detailed laser spectroscopy and Zeeman measurements. Approximately fifty-five electronic energy levels, up to the $^{4}$G$_{7/2}$ multiplet at $\sim$20 000 cm$^{-1}$ and eighty $g$ values, obtained along all three crystallographic axes, in addition to magnetic splittings determined from the previously determined $g$ tensors of the $^{6}$H$_{5/2}$Z$_{1}$ and $^{4}$G$_{5/2}$A$_{1}$ states \cite{cit:jobbitt2}, were fitted simultaneously for each site in order to obtain an unambiguous fit. Good agreement is obtained between the calculation and experimental data over the entire portion of the 4$f^{5}$ configuration for which we could make measurements.

We have also predicted the zero-field hyperfine structure for both sites of Sm$^{3+}$:Y$_{2}$SiO$_{5}$. The hyperfine splittings are estimated to be of a similar magnitude to those previously reported for Er$^{3+}$:Y$_{2}$SiO$_{5}$ \cite{horvath}. This, coupled with the small magnetic splittings of the Sm$^{3+}$ ground state suggest that this material could be a previously unexplored alternative in the development of high-bandwidth quantum information storage and communications devices.

\begin{acknowledgments}
N.L.J. would like to thank the Dodd-Walls Centre for Photonic and Quantum Technologies for the provision of a PhD studentship. The technical assistance of Mr. S. Hemmingson, Mr. R. J. Thirkettle and Mr. G. MacDonald is gratefully acknowledged.
\end{acknowledgments}

\bibliography{bibliography}

\end{document}